\documentclass[twocolumn,prb,aps,showpacs,amsmath,amssymb]{revtex4}

\usepackage{graphicx}
\usepackage{dcolumn}
\usepackage{bm}

\renewcommand{\vec}[1]{{\bf#1}}

\begin{document}

\title{Ruderman-Kittel-Kasuya-Yosida spin density oscillations: impact of the
  finite radius of the exchange interaction}

\author{Sergey Smirnov}
\affiliation{Institut f\"ur Theoretische Physik, Universit\"at Regensburg,
  D-93040 Regensburg, Germany}

\date{\today}

\begin{abstract}
A non-interacting electron gas on a one-dimensional ring is considered at
finite temperatures. The localized spin is embedded at some point on the ring
and it is assumed that the interaction between this spin and the electrons is
the exchange interaction being the basis of the Ruderman-Kittel-Kasuya-Yosida
indirect exchange effect. When the number of electrons is large enough, it
turns out that any small but finite interaction radius value can always
produce an essential change of the spin density oscillations in comparison
with the zero interaction radius traditionally used to model the
Ruderman-Kittel-Kasuya-Yosida effect.
\end{abstract}


\maketitle

\section{Introduction}
The Ruderman-Kittel-Kasuya-Yosida (RKKY) effect is a phenomenon which plays an
important role in a formation of magnetic structures in different systems. The
essence of the effect consists in an indirect interaction between localized
spins placed in a Fermi sea of non-interacting conduction electrons. The
interaction is called indirect because the spins feel the presence of each
other through the electrons surrounding them: a localized spin interacting
with the electrons induces in the electron gas spin density oscillations which
then make impact on another localized spin. The interaction underlying the
RKKY effect, that is the interaction between the localized spin and the
electron gas, can be of different nature. It can be either the hyperfine
interaction between a localized nuclear spin and conduction electrons
\cite{Rud_Kit} or the exchange interaction between the conduction and
localized electrons \cite{Kasuya,Yosida}. The latter case is realized for
example in alloys with transition metal ions where the conduction
$s$-electrons and the localized $d$-electrons of an ion interact through the
$s-d$ exchange interaction.

To model the RKKY effect it is traditionally assumed that the interaction
between a localized spin and conduction electrons is local in the real
space. This is modeled by Dirac's delta function \cite{Kittel,Harrison}. In
the case of the hyperfine interaction this model looks quite plausible because
the size of a nucleus, being of order $10^{-6}$ nm (see
Ref.~\onlinecite{Krane}), is small enough. The RKKY indirect exchange effect
based on the hyperfine interaction was studied in the scientific
literature. In Ref.~\onlinecite{Pershin} the RKKY interaction between nuclear
spins embedded in a mesoscopic ring and a finite length quantum wire was
investigated in the presence of a magnetic field. The indirect nuclear spin
interaction was found to depend on the nuclear spin positions, number of
the conduction electrons, magnetic field and system's geometry. The influence
of electron-electron interactions and electron spin correlations on the RKKY
interaction between two nuclear spins was considered by Semiromi {\it et
  al.}\cite{Semiromi} The nuclear spins were embedded in a mesoscopic metallic
ring. It was numerically shown that the electron-electron interactions and
electron spin correlations can essentially change the RKKY interaction
dependence on the magnetic flux.

However, in the case of the exchange interaction the ionic spins are much less
localized. For example, the ionic radius, which we will denote by $r_0$, for
the $f$-shell metal ions Er$^{3+}$ and Nd$^{3+}$ is $r_0=0.096$ nm and
$r_0=0.108$ nm (see Ref.~\onlinecite{Harrison_1}), respectively. The value of
$r_0$ gives an effective radius of the exchange interaction. Moreover, one can
easily conceive a situation where artificial objects like quantum dots with
non-zero total spin are embedded in a Fermi sea of conduction electrons. The
interaction between the total spin of such objects and the electrons is
similar to the exchange interaction and produces the RKKY interaction between
the total spins of those artificial objects. The value of $r_0$ can thus be
\begin{figure}
\includegraphics[width=8.5 cm]{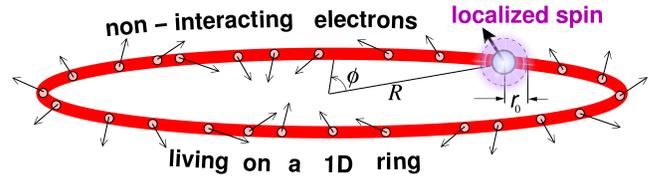}
\caption{\label{figure_1} (Color online) A non-interacting electron gas on a
  1D ring of radius $R$. The localized spin is centered around the ring point
  with the polar angle $\phi=0$. The localization radius of the spin is $r_0$
  which is also estimated to be the radius of the exchange interaction in the
  RKKY effect.}
\end{figure}
technologically varied. Systems with quantum dots interacting through the
RKKY effect were already investigated in a number of scientific papers. The
RKKY effect between two quantum dots embedded in an Aharonov-Bohm ring was
investigated by Utsumi {\it et al.}\cite{Utsumi} as a function of a magnetic
flux through the ring. The quantum dots contained odd numbers of
electrons. The interaction between the total quantum dots' spins and the
conduction electrons in the ring was described by a tunneling Hamiltonian. In
this tunneling Hamiltonian the coupling constant was modeled by the delta
function that is the interaction radius was $r_0=0$. In
Ref.~\onlinecite{Rikitake} two localized spins in one-, two- and
three-dimensional electron gases were considered. Decoherence of the spins was
studied using the kinetic equation for the reduced density
matrix. Additionally, a quantum gate consisting of two quantum dots embedded
in a two-dimensional electron gas of GaAs/AlGaAs heterostructure was
investigated. The RKKY effect was provided through the $s-d$ exchange
interaction which was assumed to take place just at the positions of the
localized spins, that is the interaction radius was zero. Tamura {\it et
  al.}\cite{Tamura} studied the RKKY interaction between the localized spins
of two quantum dots placed at the opposite edges of a one-dimensional (1D)
conducting channel. The RKKY interaction between the spins across the channel
was taken into account by virtue of an exchange interaction where the Fourier
transform of the coupling constant was assumed to be momentum-dependent. This
in principle means that the interaction could be non-local. However,
consequences of this non-locality were out of focus of that work. In
Ref.~\onlinecite{Aono} the RKKY interaction in a coupled quantum dot system
embedded in a ring with a spin-orbit interaction was explored in the presence
of the Aharonov-Bohm and Aharonov-Casher effects. The $s-d$ exchange
interaction responsible for the RKKY effect in this system was also local.
\begin{figure}
\includegraphics[width=8.5 cm]{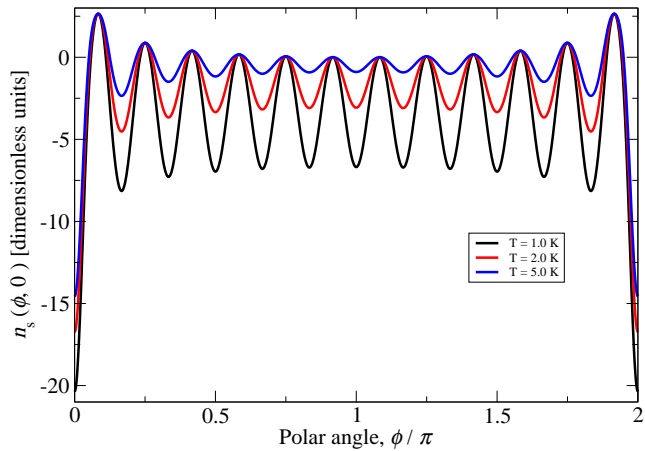}
\caption{\label{figure_2} (Color online) The function $n_\mathrm{s}(\phi,r_0)$
  describing the spin density oscillations on a mesoscopic ring for the
  conventional model with $r_0=0$. The number of electrons is
  $N_\mathrm{el}=24$.}
\end{figure}

Finally, we would like to mention that effects of the finite size of the ionic
spin distribution on the RKKY interaction have been studied in bulk systems in
connection with ferromagnetic Heusler alloys
\cite{Darby,Smit,Darby_1,Malmstroem,Smit_1}. However, the models used in those
attempts did not give the delta function in the limit $r_0\rightarrow 0$.

The purpose of the present work is to verify by means of a simple model
whether in a mesoscopic ring a finite value of $r_0$ can play any role in the
formation of the electron spin density oscillations which underpin the RKKY
indirect exchange interaction. In the limit $r_0\rightarrow 0$ the model which
we use gives the delta function. It is shown that if the interaction radius is
finite, its impact on the electron spin density oscillations is to reduce the
oscillation amplitude. An interesting feature is that when the number of the
electrons in the gas is large enough, this reduction, produced by the presence
of a small domain of size $r_0$, is significant and takes place in the whole
system with much larger size.

The paper is organized as follows. Sec. \ref{TM} describes a mathematical
formulation of the problem. In Sec. \ref{ITGFS} the solution is obtained using
the Matsubara diagrammatic approach. The results are discussed in
Sec. \ref{RD}.

\section{Theoretical model}\label{TM}
We consider a non-interacting Fermi-gas on a 1D ring of radius $R$. The
electron positions are specified by the polar angle $\phi$. An ion (or another
object) with a non-zero total spin is placed on the ring at $\phi=0$. The
system is schematically shown in Fig.~\ref{figure_1}. The interaction between
the localized spin and conduction electrons is assumed to be an exchange
interaction which in general can be non-local.

To formulate the problem mathematically we write down the Hamiltonian of the
system in the form:
\begin{equation}
\hat{H}=\hat{H}_0+\hat{H}_\mathrm{int}.
\label{Ham}
\end{equation}
In Eq. (\ref{Ham}) $\hat{H}_0$ is the Hamiltonian of the non-interacting
$N$-electron system on a 1D ring:
\begin{equation}
\hat{H}_0=\sum_{i=1}^N \hat{H}_0^i,
\label{Ham_0}
\end{equation}
with
\begin{equation}
\hat{H}_0^i=-\frac{\hbar^2}{2mR^2}\frac{\partial^2}{\partial\phi_i^2}
\label{Ham_0_i}
\end{equation}
where $m$ is the electron mass and $\phi_i$ is the $i$-th electron
coordinate. The spin-degenerate eigen-values of the single-particle
Hamiltonian $\hat{H}_0^i$ are
\begin{equation}
\epsilon_{n\sigma}=\frac{\hbar^2n^2}{2mR^2},
\label{Ham_0_i_ev}
\end{equation}
where $n=0,\pm 1,\ldots$ and $\sigma$ is the spin index. The normalized
eigen-states of $\hat{H}_0^i$, $|n\sigma\rangle$, in the coordinate
representation are
\begin{equation}
\langle\phi\sigma'|n\sigma\rangle=\delta_{\sigma\sigma'}\frac{1}{\sqrt{2\pi}}\exp(-in\phi).
\label{Ham_0_i_es}
\end{equation}
The term $\hat{H}_\mathrm{int}$ in Eq. (\ref{Ham}) describes the exchange
interaction between the localized spin and the electron gas and it is
conventionally written as:
\begin{equation}
\hat{H}_\mathrm{int}=\sum_{j=1}^N J(\phi_j)S^i\sigma_j^i,
\label{Ham_int}
\end{equation}
where $J(\phi)$ is the coupling function of the exchange interaction,
$\vec{S}$ is the localized spin placed at $\phi=0$, $\vec{\sigma}_j$ is the
vector of the electron spin Pauli matrices and the summation over the index
$i$ is assumed.

As it was mentioned above, traditionally it is assumed that the exchange
interaction is local, that is the embedded spin interacts with the electrons
only at $\phi=0$. In this case the coupling function $J(\phi)$ is modeled by
the following dependence on the polar angle:
\begin{equation}
J(\phi)=J\delta(\phi),
\label{J_loc}
\end{equation}
where $J$ is a coupling constant.

In this work we suggest a simple model in which the localized spin interacts
with the electrons in a small vicinity of the point $\phi=0$ and when the
vicinity is tightened into a point at $\phi=0$, the model turns into the
conventional one (\ref{J_loc}):
\begin{equation}
J(\phi)=\frac{J}{\phi_0\sqrt{\pi}}
\begin{cases}
\exp\bigl[-\bigl(\frac{\phi}{\phi_0}\bigl)^2\bigl],&
0\leqslant\phi<\pi,\\
\exp\bigl[-\bigl(\frac{2\pi-\phi}{\phi_0}\bigl)^2\bigl],&
\pi\leqslant\phi<2\pi.
\end{cases}
\label{J_non_loc}
\end{equation}
The size $\phi_0=r_0/R$ of this vicinity is estimated from the radius $r_0$ of
the ion (or from a characteristic size of another object) which is the source
of the spin centered around $\phi=0$. Using the well known representation of
Dirac's delta function,
\begin{equation}
\delta(x)=\underset{t\rightarrow
  0}{\mathrm{lim}}\frac{\exp\bigl[-\bigl(\frac{x}{\sqrt{t}}\bigl)^2\bigl]}{\sqrt{\pi t}},
\label{df_rep}
\end{equation}
one is convinced that the non-local model, Eq. (\ref{J_non_loc}), takes the
local form, Eq. (\ref{J_loc}), when $r_0\rightarrow 0$.
\begin{figure}
\includegraphics[width=8.5 cm]{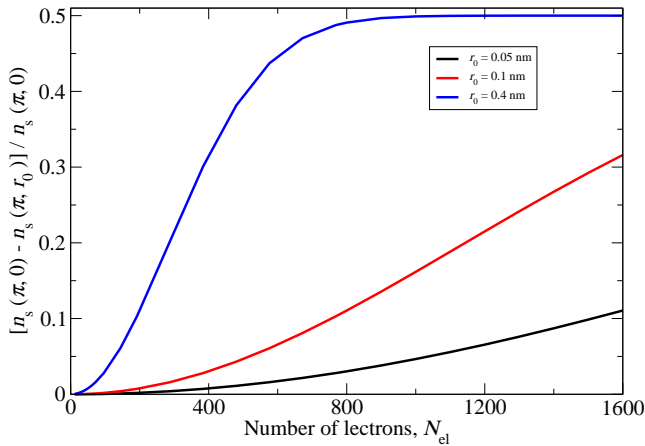}
\caption{\label{figure_3} (Color online) The relative change of the spin
  density at $\phi=\pi$ (the furthermost point from the localized spin). The
  reference is the spin density at $r_0=0$. The temperature is $T=1 K$.}
\end{figure}

To study whether finite values of $r_0$ have any effect we will calculate the
electron spin density at finite temperatures. The electron spin density
operator at point $\phi$ on the ring is defined as
\begin{equation}
\hat{n}_\mathrm{s}^i(\phi)=\sum_{j=1}^{N}\delta(\phi_j-\phi)\sigma_j^i.
\label{sd_op}
\end{equation}
The electron spin density at point $\phi$ is obtained through the statistical
average using the Gibbs grand canonical ensemble:
\begin{equation}
n_\mathrm{s}^i(\phi,r_0)=\frac{\mathrm{Tr}
\bigl[\exp\bigl(-\frac{\hat{H}-\mu\hat{N}}{k_\mathrm{B}T}\bigl)\hat{n}_s^i(\phi)\bigl]}{Z},
\label{sd}
\end{equation}
where we explicitly show that the electron spin density is also a function of
the interaction radius $r_0$. In Eq. (\ref{sd}) $Z$ is the partition function:
\begin{equation}
Z=\mathrm{Tr}\biggl[\exp\biggl(-\frac{\hat{H}-\mu\hat{N}}{k_\mathrm{B}T}\biggl)\biggl].
\label{pf}
\end{equation}
In Eqs. (\ref{sd}) and (\ref{pf}) $k_\mathrm{B}$ is the Boltzmann constant,
$\mu$ is the chemical potential, $T$ is the temperature, $\hat{N}$ is the
particle number operator and the trace is taken using a complete set of states
of the Fock space. It is also important to note that since the number $N$ of
electrons is fixed, the chemical potential $\mu$ is not an independent
variable but a function of the temperature $T$.

\section{Imaginary time Green's function solution}\label{ITGFS}
The problem formulated in the previous section is obviously not solvable
exactly. Thus some approximation methods should be applied. For small values
of the coupling constant $J$ in Eq. (\ref{J_non_loc}) a perturbation theory,
where $\hat{H}_\mathrm{int}$ is considered as a perturbation, can be
applied. As the calculations are performed at finite temperatures, instead of
the pure quantum mechanical perturbation theory one has to use the so-called
thermodynamic perturbation theory \cite{Landau_V} for the quantum statistical
averages. This perturbation theory is in general valid when the perturbation
energy per particle is less than $k_\mathrm{B}T$, i.e., $J<k_\mathrm{B}T$.
However, very often it happens that when $T\rightarrow 0$, the coefficients of
the perturbation expansion change as functions of $T$ in such a way that the
thermodynamic perturbation theory can remain valid for all temperatures.

Although the thermodynamic perturbation theory gives a general approach to
calculate statistical averages, in its original form it is quite
cumbersome. It is more convenient to use this theory reformulated in terms of
a diagrammatic approach, namely the Matsubara (or imaginary time) diagrammatic
method \cite{AGD}.

In order to employ this technique for our purposes we first rewrite the total
Hamiltonian of the problem, Eq. (\ref{Ham}), in the second quantized form
using the $\{n\sigma\}$ single-particle basis:
\begin{equation}
\begin{split}
&\hat{H}=\sum_{n\sigma}\epsilon_{n\sigma}a^\dagger_{n\sigma}a_{n\sigma}+\frac{JS^i}{2\pi}
\sum_{n\sigma\,n'\sigma'}\langle\sigma|\hat{\sigma}^i|\sigma'\rangle\times\\
&\times\biggl\{\frac{1}{\phi_0\sqrt{\pi}}\int_\pi^\pi\,d\tilde{\phi}
\exp\biggl[-\biggl(\frac{\tilde{\phi}}{\phi_0}\biggl)^2\biggl]\times\\
&\times\exp[i\tilde{\phi}(n-n')]\biggl\}a^\dagger_{n\sigma}a_{n'\sigma'}.
\end{split}
\label{Ham_sq}
\end{equation}
The second quantized form of the electron spin density operator at point
$\phi$ is
\begin{equation}
\hat{n}_\mathrm{s}^i(\phi)=\frac{1}{2\pi}
\sum_{n\sigma\,n'\sigma'}
\langle\sigma|\hat{\sigma}^i|\sigma'\rangle\exp[i\phi(n-n')]a^\dagger_{n\sigma}a_{n'\sigma'}
\label{sd_op_sq}
\end{equation}
and the expression for the electron spin density at point $\phi$ may be
rewritten as
\begin{equation}
n_\mathrm{s}^i(\phi,r_0)=-\sum_{\sigma\sigma'}
\langle\sigma'|\hat{\sigma}^i|\sigma\rangle\mathcal{G}_{\sigma\sigma'}(\phi,\tau;\phi,\tau).
\label{sd_Gf}
\end{equation}
In the last expression $\mathcal{G}_{\sigma\sigma'}(\phi,\tau;\phi',\tau')$ is
the one-particle imaginary time Green's function defined as
\begin{equation}
\begin{split}
&\mathcal{G}_{\sigma\sigma'}(\phi,\tau;\phi',\tau')=\\
&=\frac{1}{Z}\mathrm{Tr}\biggl[\exp\biggl(-\frac{\hat{H}-\mu\hat{N}}{k_\mathrm{B}T}\biggl)
\mathrm{T}\hat{\psi}_{\sigma}(\phi,\tau)\hat{\psi}_{\sigma'}^\dagger(\phi',\tau')\biggl],
\end{split}
\label{op_it_Gf}
\end{equation}
where $\mathrm{T}$ is the time-ordering operator and
$\hat{\psi}_{\sigma}(\phi,\tau)$ are the imaginary time field operators,
\begin{equation}
\begin{split}
&\hat{\psi}_{\sigma}(\phi,\tau)=\\
&=\exp[\tau(\hat{H}-\mu\hat{N})]\hat{\psi}_\sigma(\phi)\exp[-\tau(\hat{H}-\mu\hat{N})],
\end{split}
\label{it_fo}
\end{equation}
with $\hat{\psi}_\sigma(\phi)$ related to the annihilation operators
$a_{n\sigma}$ as
\begin{equation}
\hat{\psi}_\sigma(\phi)=\frac{1}{\sqrt{2\pi}}\sum_n\exp(-i\phi n)a_{n\sigma}.
\label{fo_ao}
\end{equation}
We now apply the diagrammatic expansion of the Green's function
$\mathcal{G}_{\sigma\sigma'}(\phi,\tau;\phi',\tau')$. The effect of the RKKY
spin density oscillations appears already in the first order and thus we only
need to consider the first order diagram. Such an approach to the RKKY spin
density oscillations was considered in Ref.~\onlinecite{LS} for a
three-dimensional electron gas. However, in that case the electron spectrum
was continuous and to perform momentum integrals the linearization of the
spectrum on the Fermi-surface was employed to get the long range behavior of
the RKKY oscillations. In our problem the electron spectrum is discrete and
instead of integrals we will have sums over discrete indices. Moreover, since
our system is finite we are interested in the RKKY oscillations in the whole
range and not only at long distances from the localized spin.
\begin{figure}
\includegraphics[width=8.5 cm]{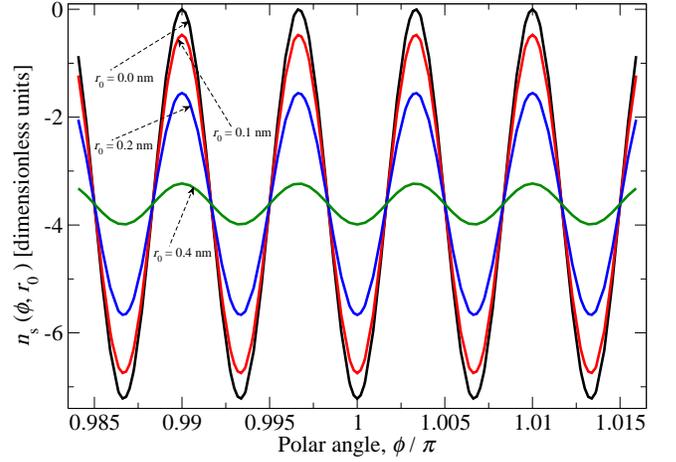}
\caption{\label{figure_4} (Color online) The electron spin density
  oscillations in the vicinity of the point $\phi=\pi$. The number of
  electrons is $N_\mathrm{el}=600$. The temperature is $T=1 K$.}
\end{figure}

The first order contribution to the Green's function
$\mathcal{G}_{\sigma\sigma'}(\phi,\tau;\phi',\tau)$ is
\begin{equation}
\begin{split}
&\mathcal{G}_{\sigma\sigma'}^{(1)}(\phi,\tau;\phi',\tau)=-\frac{JS^i}{(2\pi)^2\hbar}
\langle\sigma|\hat{\sigma}^i|\sigma'\rangle\times\\
&\times\sum_{n\,n'}\exp(-in\phi+in'\phi')j_{n\,n'}s_{n\,n'},
\end{split}
\label{Gf_1o}
\end{equation}
where
\begin{equation}
\begin{split}
&j_{n\,n'}=\exp\biggl[-\frac{1}{4}(n-n')^2\phi_0^2\biggl]\times\\
&\times\mathrm{Re}
\biggl\{\mathrm{erf}\biggl[\frac{\pi}{\phi_0}+\frac{i}{2}(n-n')\phi_0\biggl]\biggl\},
\end{split}
\label{j_f}
\end{equation}
and
\begin{equation}
s_{n\,n'}=
\begin{cases}
\hbar\frac{n_{n'}-n_n}{\epsilon_{n'}-\epsilon_{n}},& n'^2\neq n^2,\\
-n_n(1-n_n)\frac{\hbar}{k_\mathrm{B}T},& n'^2=n^2
\end{cases}
\label{s_f}
\end{equation}
with $n_n$ being the fermion occupation numbers,
\begin{equation}
n_n=\frac{1}{\exp(\frac{\epsilon_n-\mu}{k_\mathrm{B}T})+1}.
\label{fon}
\end{equation}
Since $\phi_0\ll\pi$, we have
\begin{equation}
j_{n\,n'}\approx\exp\biggl[-\frac{1}{4}(n-n')^2\phi_0^2\biggl]
\label{j_approx}
\end{equation}
and this approximation will be used in the calculations below.

Substituting Eq. (\ref{Gf_1o}) into Eq. (\ref{sd_Gf}) and taking into account
Eq. (\ref{s_f}), we obtain the first order contribution to the electron spin
density
\begin{equation}
n_\mathrm{s}^i(\phi,r_0)=\frac{2JS^i}{(2\pi)^2}n_\mathrm{s}(\phi,r_0),
\label{sd_fe}
\end{equation}
where $n_\mathrm{s}(\phi,r_0)$ is given by the following expression:
\begin{equation}
\begin{split}
&n_\mathrm{s}(\phi,r_0)=
\!\!\!\!\!\sum_{\substack{n\,n'\\(n'^2\neq n^2)}}\exp[-i(n-n')\phi]j_{n\,n'}
\frac{n_{n'}-n_n}{\epsilon_{n'}-\epsilon_n}-\\
&-\sum_{\substack{n\,n'\\(n'^2=n^2)}}
\exp[-i(n-n')\phi]j_{n\,n'}\frac{n_n(1-n_n)}{k_\mathrm{B}T}.
\end{split}
\label{sd_ad}
\end{equation}
For a given value of the interaction radius $r_0$ the function
$n_\mathrm{s}(\phi,r_0)$ provides an oscillatory behavior of the electron spin
density as a function of the polar angle $\phi$.

\section{Results and discussion}\label{RD}
In this section we numerically analyze the electron spin density
oscillations. Using Eq. (\ref{sd_ad}) we calculate the function
$n_\mathrm{s}(\phi,r_0)$. The mesoscopic ring is assumed to be fabricated on
AlGaAs-GaAs heterostructures. The values of the parameters are taken close to
the ones used in previous works \cite{Utsumi} and in experiments
\cite{Fuhrer}. The ring radius is $R=40$ nm, the effective mass $m=0.067m_0$,
where $m_0$ is the free electron mass.

Let us first consider the behavior of $n_\mathrm{s}(\phi,r_0)$ as a function
of the polar angle for the conventional model with $r_0=0$. It is shown in
Fig.~\ref{figure_2} for different values of the temperature $T$ and for the
number of electrons $N_\mathrm{el}=24$. In systems with a continuous spectrum
the RKKY spin density oscillations behave like $\cos(2p_\mathrm{F}r/\hbar)$
(similar to the Friedel oscillations of the electronic density) where
$p_\mathrm{F}$ is the Fermi momentum and $r$ is the distance from the
localized spin. In our case the spectrum is discrete. The analog of the Fermi
momentum is the energy level number $n_\mathrm{F}$ such that at $T=0$ the
states with $|n|>n_\mathrm{F}$ are not populated. Since the state with a given
value of $n$ can be occupied by two electrons, for $N_\mathrm{el}=24$ one gets
$n_\mathrm{F}=6$ and thus $2n_\mathrm{F}=12$. In agreement with this estimation
Fig.~\ref{figure_2} shows 12 oscillations. As it was discussed in the
literature (see, for example, Ref.~\onlinecite{LS}), the effect of the
temperature on the RKKY spin density oscillations is to produce a faster
reduction of the oscillation amplitude with the distance from the localized
spin. For example in three-dimensional electron gases at $T=0$ the oscillation
amplitude at large distances decreases as $1/r^3$ and at finite temperatures
it is damped at the thermal distance $\hbar p_\mathrm{F}/(2\pi mk_\mathrm{B}T)$.
An analogous behavior takes place in our case as well. As it can be seen from
Fig.~\ref{figure_2} the amplitude of the oscillations as a function of the
polar angle is damped faster for higher values of the temperature.

Next we turn our attention to the effect of the finite interaction radius
$r_0$ of the exchange interaction on the electron spin density
oscillations. From Eq. (\ref{j_approx}) it seems that for realistic, that is
small, values of $\phi_0$ the function $j_{n\,n'}\approx 1$ and a finite
interaction radius value does not produce any change in the electron spin
density oscillations in comparison with the case $r_0=0$. However, this
reasoning is not entirely correct because it does not take into consideration
the number of electrons in the gas surrounding the localized spin in the
ring. Indeed, when the number of electrons in the gas grows, energy levels
with higher values of $|n|$ become important. The contributions with higher
values of $|n-n'|$ are involved. From Eqs. (\ref{j_approx}) and (\ref{sd_ad})
one observes that in parallel with this involvement the contributions from
terms with high values of $|n-n'|$ get more suppressed. It does not matter how
small the interaction radius is. The main point is that it is finite. For any
finite value of $r_0$ there exists a number of electrons $N_\mathrm{c}$ in the
gas such that for $N_\mathrm{el}>N_\mathrm{c}$ the interaction radius, no
matter how small it is, will produce more and more pronounced impact on the
spin density. This is demonstrated in Fig.~\ref{figure_3} where the relative
change of the spin density at $\phi=\pi$ is depicted as a function of the
number of the electrons in the gas for different values of the interaction
radius $r_0$. As expected the change of the electron spin density in
comparison with the conventional model with $r_0=0$ is negligible for small
numbers of the electron in the ring. For larger numbers of the electrons the
electron spin density for finite values of $r_0$ starts to deviate from the
model with $r_0=0$. It is interesting that even the interaction radius
$r_0=0.05$ nm can produce an observable change of approximately 11\% for
$N_\mathrm{el}=1600$. The ring distance between the two points $\phi=0$ and
$\phi=\pi$ is about 126 nm while the size of the area where the spin is
localized is 0.1 nm, that is three orders of magnitude less. An important
result is that the small vicinity around the localized spin is able to
significantly change the electron spin density in every point of the system
whose size is several orders of magnitude larger than the size of the domain
over which the localized spin is spread.

The RKKY spin density oscillations for $N_\mathrm{el}=600$ are displayed in
Fig.~\ref{figure_4} for different values of $r_0$. In this case the number
of the oscillations is approximately equal to 300 and thus only a small
vicinity around $\phi=\pi$ is plotted to clearly show the oscillations.

Finally, we note that semi-quantitatively the oscillating behavior can be
explained by the dominance of the term $\cos(2n_\mathrm{F}\phi)
j_{n_\mathrm{F},-n_\mathrm{F}}n_{n_\mathrm{F}}(1-n_{n_\mathrm{F}})$ in Eq. (\ref{sd_ad}).
The $2n_\mathrm{F}$ RKKY oscillations with $r_0=0$ are weighted with
$j_{n_\mathrm{F},-n_\mathrm{F}}<1$. For small $N_\mathrm{el}$ the weight
$j_{n_\mathrm{F},-n_\mathrm{F}}\approx 1$ and plays no role but for large
$N_\mathrm{el}$ this weight reduces the amplitude of the $2n_\mathrm{F}$
RKKY oscillations by the factor $\exp[-(n_\mathrm{F}\phi_0)^2]$.

\section{Summary}\label{S}
The Ruderman-Kittel-Kasuya-Yosida (RKKY) spin density oscillations, induced by
an exchange interaction between a localized spin and the electron gas in which
the spin is embedded, have been investigated at finite temperatures taking
into account finite values of the exchange interaction radius. It has been
found that the electron spin density in a non-interacting gas on a mesoscopic
ring is changed in comparison with the traditional model which assumes that
the interaction radius is zero. The amplitude of the RKKY oscillations is
suppressed when the number of the electrons in the gas increases. This
suppression gets stronger for larger values of the interaction radius.

A remarkable point is that as soon as the interaction radius is finite, it is
already not important how small it is because its impact always becomes
observable when the number of the electrons in the ring is large enough.

\begin{acknowledgments}
The author thanks Prof. Milena Grifoni for useful discussions and
comments. Support from the DFG under the program SFB 689 is acknowledged.
\end{acknowledgments}

\end{document}